\documentclass[a4paper,11pt]{article}
\usepackage{pos}

\newcommand{\qhat}{\hat{q}}
\newcommand{\dpi}{(2\pi)}
\newcommand{\calO}{\mathcal{O}}
\newcommand{\Msq}{|\bar{\mathcal{M}}|^2}
\renewcommand{\vec}{\mathbf}

\renewcommand{\logo}{\relax}
\graphicspath{ {figures/} }

\title{Exploring jet transport coefficients by elastic and radiative scatterings in the strongly interacting quark-gluon plasma}
\ShortTitle{Exploring jet transport coefficients in the QGP}

\author*[a]{Ilia Grishmanovskii}
\author[b]{Taesoo Song}
\author[a,c]{Olga Soloveva}
\author[a,c]{Carsten Greiner}
\author[a,b,c]{Elena Bratkovskaya}

\affiliation[a]{Institut f\"ur Theoretische Physik, Johann Wolfgang Goethe-Universit\"at,\\ Max-von-Laue-Str.\ 1, Frankfurt am Main, Germany}

\affiliation[b]{GSI Helmholtzzentrum f\"{u}r Schwerionenforschung GmbH,\\ Planckstrasse 1, Darmstadt, Germany}

\affiliation[c]{Helmholtz Research Academy Hesse for FAIR (HFHF), GSI Helmholtz Center for Heavy Ion Physics,\\ Ruth-Moufang-Straße 1, Frankfurt am Main, Germany}

\emailAdd{grishm@itp.uni-frankfurt.de}
\emailAdd{soloveva@itp.uni-frankfurt.de}
\emailAdd{T.Song@gsi.de}
\emailAdd{E.Bratkovskaya@gsi.de}

\abstract{We investigate the interaction of leading jet partons  within a strongly interacting quark-gluon plasma (sQGP) medium, using the effective dynamical quasiparticle model (DQPM). The DQPM offers a description of the sQGP's non-perturbative nature at finite temperature $T$ and baryon chemical potential $\mu_B$ through a propagator representation of massive off-shell partons (quarks and gluons). These partons are characterized by spectral functions with $T,\mu_B$ dependent masses and widths, adjusted to reproduce the lattice Quantum Chromodynamics (lQCD) equation-of-state (EoS) for the QGP in thermodynamic equilibrium. Our focus lies on examining the jet transport coefficients by elastic scattering in sQGP, specifically the transverse momentum transfer squared per unit length denoted as $\hat{q}$, within the QGP. Furthermore, we investigate the dependence of these coefficients on both the medium temperature $T$ and the jet parton energy. By studying the jet transport coefficients and their relationship to temperature and parton energy, we aim to gain insights into the dynamics of jet propagation in the strongly interacting quark-gluon plasma medium.}

\FullConference{HardProbes2023\\
 26-31 March 2023\\
 Aschaffenburg, Germany\\}

\begin{document}
\maketitle


Ultra-relativistic heavy-ion collisions performed at the Super Proton Synchrotron (SPS), the Relativistic Heavy-Ion Collider (RHIC) and the Large Hadron Collider (LHC) at CERN provide an access to a new hot and dense phase of matter, the quark-gluon plasma (QGP). An understanding of the properties of the QGP is one of the main goals of current research in heavy-ion physics.

Jet quenching appeared to be an effective tool for investigating the properties of the quark-gluon plasma (QGP) matter. Produced in the early stage of the heavy-ion collisions, jets get high transverse momentum and traverse the QGP interacting with the medium through collisional and radiative processes. Starting from the pioneering work by Bjorken in 1982~\cite{Bjorken:1982} a substantial progress in the understanding of the jet energy loss has been made in the last decades. Theoretical studies \cite{Thoma:1990fm} showed that the parton energy loss can be described by a series of jet transport coefficients such as the jet quenching parameter $\qhat$ (the transverse momentum transfer squared per unit length of the propagating hard parton) or energy loss coefficient $\Delta E = dE/dx$ (energy loss per unit length).

In a recent publication (Ref. \cite{Grishmanovskii:2022tpb}), we investigated the jet transport coefficients $\qhat$ and $\Delta E$ using the dynamical quasiparticle model (DQPM) as an effective field-theoretical model \cite{Peshier:2005pp, Berrehrah:2016vzw, Moreau:2019vhw}. The focus of our study was on the energy loss of jet partons primarily due to elastic scattering, which is expected to dominate at low and intermediate jet momenta. In this contribution, we will review the key findings regarding the $\qhat$ coefficient as reported in Ref. \cite{Grishmanovskii:2022tpb}.


\section{Transport coefficients in DQPM}

The DQPM \cite{Peshier:2005pp, Berrehrah:2016vzw, Moreau:2019vhw} provides an effective approach to describe the quark-gluon plasma (QGP) in terms of strongly interacting quarks and gluons. Their properties are fitted to match lattice QCD calculations in thermal equilibrium and at zero quark chemical potential. In the DQPM, quasiparticles are characterized by dressed propagators with complex self-energies, where the real part of the self-energies represents dynamically generated thermal masses, while the imaginary part carries information about the partons reaction rates (which is twice the widths, i.e. inverted lifetimes). The off-shell partonic interaction cross sections in the DQPM are evaluated based on leading-order scattering diagrams and depend on the temperature ($T$), baryon chemical potential ($\mu_B$), the invariant energy of the colliding partons ($\sqrt{s}$), and the scattering angle \cite{Moreau:2019vhw}.

The general expression for a transport coefficient in kinetic theory, accounting for off-shell medium partons, is given by the following expression:
\begin{align}
  \langle \calO \rangle^{\text{off}} &= \frac{1}{2E_i}\sum_{j=q,\bar{q},g}
  \int\frac{d^4p_j}{\dpi^4} d_j f_j \tilde{\rho}(\omega_j,\vec{p}_j) \theta(\omega_j)
  \int\frac{d^3p_1}{\dpi^3 2E_1} \int\frac{d^4p_2}{\dpi^4} \tilde{\rho}(\omega_2,\vec{p}_2)\theta(\omega_2) 
  \nonumber\\
  &\times (1 \pm f_1)(1 \pm f_2)\ \calO \ 
  \Msq\ (2\pi)^4 \delta^{(4)}(p_i + p_j - p_1 - p_2).
  \label{eq:O_off}
\end{align}
In this expression, $d_j$ represents the degeneracy factor for spin and color, $\tilde{\rho}(\omega_i)$ are renormalized spectral functions, and $f_j$ are the Fermi (for quarks) or Bose (for gluons) distribution functions. The Pauli-blocking (-) and Bose-enhancement (+) factors account for the available density of final states. The sum denoted by $\sum_{j=q,\bar{q},g}$ includes the contribution from all possible partons, which in this specific study consist of gluons and (anti-) quarks of three different flavors ($u, d, s$).

Depending on the choice of $\calO$ in equation \eqref{eq:O_off} one can refer to different transport coefficients:
\begin{itemize}
    \item $\calO = |\vec{p_T} - \vec{p_T^{\prime}}|^2$ -- to the jet transport coefficient $\qhat$,
    \item $\calO = (E-E')$ -- to the energy loss $\Delta E = dE/dx$,
\end{itemize}
where $p_T$ represents the transverse momentum.


\section{Results}

\begin{figure}[b!]
  \centering
  \includegraphics[width=0.5\textwidth]{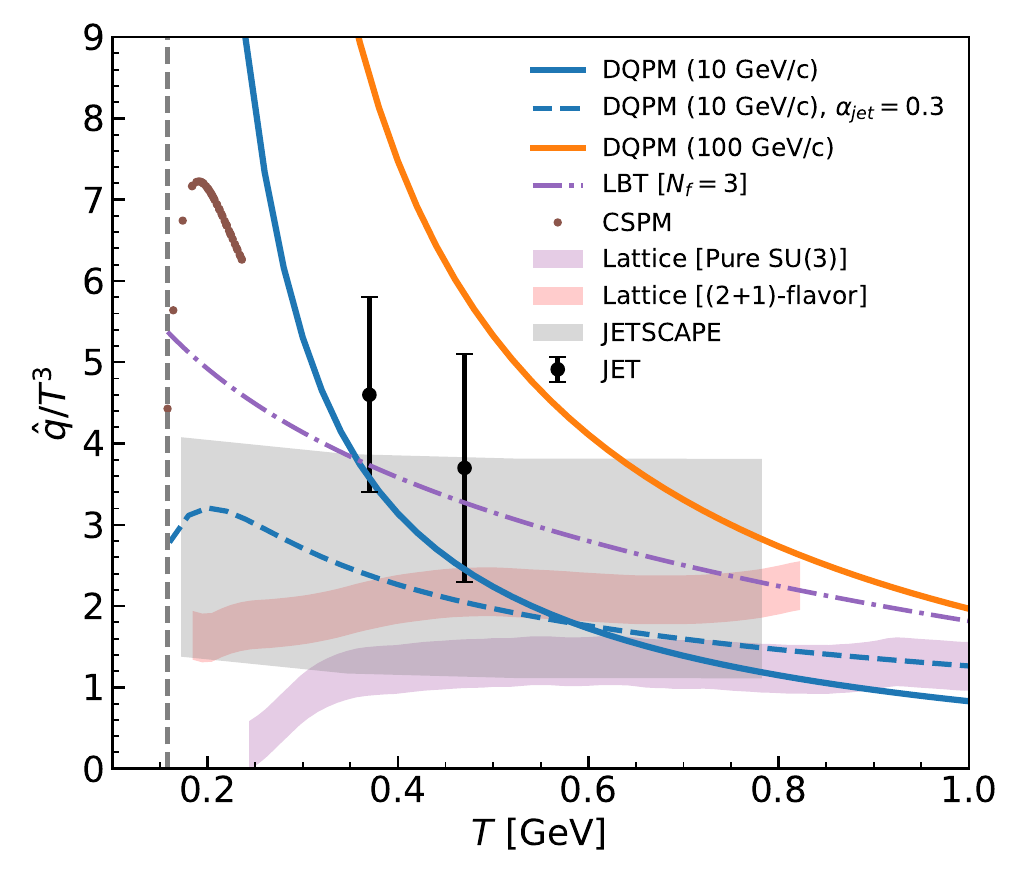}
  \caption{
  Temperature dependence of the scaled jet transport coefficient $\hat{q}/T^3$. The off-shell DQPM results are represented for a quark jet with mass $M=0.01$~GeV and momentum 10 GeV/c (blue line) and 100 GeV/c (orange line). The blue dashed line shows the DQPM result with $\alpha_{S}^{jet}=0.3$ at the jet parton vertices.
  The purple dash-dotted line represents the LBT results for $N_f=3$ and $p=10$ GeV/c \cite{He:2015pra}, while the red and purple areas represent lQCD estimates \cite{kumar2020jet} for pure SU(3) gauge theory and (2+1) flavour QCD, respectively, in the limit of an infinitely hard jet parton. The gray area corresponds to the results from the JETSCAPE Collaboration ($p = 100$ GeV/c) \cite{JETSCAPE:2021ehl}. The black dots represent the phenomenological extraction by the JET Collaboration presented for $p=10$ GeV/c \cite{Burke:2013yra}, while the brown dots show the results from the color string percolation model (CSPM) \cite{Mishra:2021yer}. The vertical gray dashed line indicates the critical temperature $T_C=0.158$ GeV.
  The figure is taken from Ref. \cite{Grishmanovskii:2022tpb}.}
  \label{fig:qhat-T}
\end{figure}

In Fig. \ref{fig:qhat-T} we show the temperature dependence of the scaled $\qhat/T^3$ transport coefficient for $\mu_B=0$ for different approaches. The DQPM results are obtained by full off-shell calculations. In general, there are four effects that lead to the DQPM results being different from pure pQCD calculations:\\
$\bullet$ Strong coupling is dominantly responsible for the sensitivity to the transport properties of the QCD medium since it enters the definitions of thermal masses/widths and scattering amplitudes. The strong temperature dependence of the coupling leads to a strong temperature dependence of the transport coefficients.\\
$\bullet$ The finite masses of intermediate parton propagators play the same role as the Debye screening mass in HTL calculations, providing the cut-off effect and a general suppression for the differential cross sections.\\
$\bullet$ The finite masses of the medium partons have three effects on the total value of the transport coefficients:
Firstly, they enter the expression for the scattering amplitude and have a large effect for small scattering energies. For high energies, however, the effect of finite masses becomes negligible.
Secondly, parton masses enter the definition of transport coefficients, which leads to an increase in $\qhat$.
Thirdly, parton masses enter the distribution function of thermal partons $f(E,T,\mu_B)$, leading to a strong suppression of the transport coefficients.
This effect is dominant.\\
$\bullet$ The finite widths of partons also have a small effect on the scattering amplitudes, but they are important for the off-shell calculations as they define the shape of the spectral function.
Thus, eventually, a large sensitivity of the jet energy loss to the properties of the QCD medium comes not only from the strong coupling but from all aspects of the DQPM model.\\
As follows from Fig. \ref{fig:qhat-T}, there are large model uncertainties in the determination of $\qhat$ from both theoretical and phenomenological sides.

\begin{figure}[t!]
  \centering
  \includegraphics[width=0.8\textwidth]{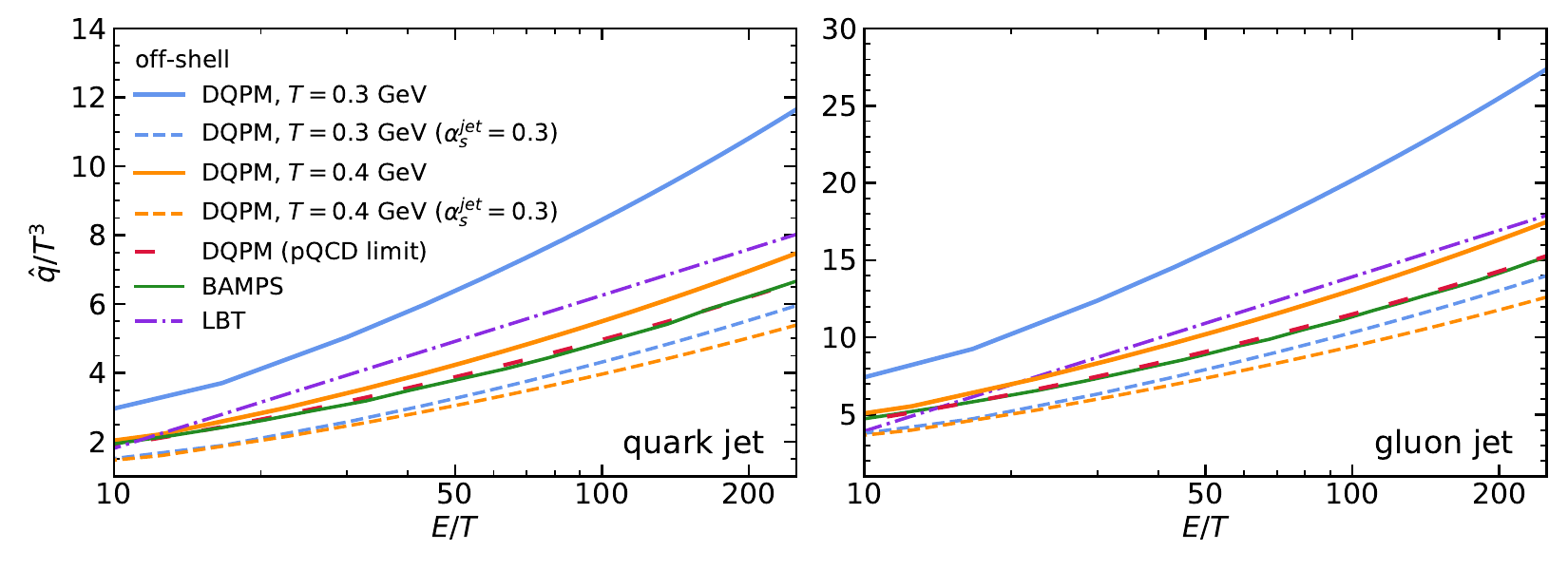}
  \caption{The scaled $\qhat/T^3$ coefficients as a function of $E/T$ for a quark jet (left) and a gluon jet (right) for different medium temperatures. The blue solid (upper) and orange solid (middle) lines represent the off-shell DQPM result for $T=0.3$ and $0.4$ GeV, respectively. The dashed lines of the same color represent DQPM results with $\alpha_{jet}=0.3$ at jet parton vertices for $T=0.3$ and $0.4$ GeV, respectively. The red long-dashed line represents the DQPM result in the pQCD limit. The green solid line represents BAMPS results \cite{Senzel:2020tgf} and the purple dash-dotted line stands for the LBT model results \cite{He:2015pra}. All calculations include only elastic energy loss.
  The figure is taken from Ref. \cite{Grishmanovskii:2022tpb}.
  }
  \label{fig:qhat_p}
\end{figure}

In Figure \ref{fig:qhat_p}, we present the scaled $\qhat/T^3$ coefficients for both a quark (left) and a gluon jet (right) resulting from elastic scattering with off-shell medium partons (from Eq. \eqref{eq:O_off}). These coefficients are plotted as a function of the ratio of the jet energy to temperature, denoted as $E/T$. It is observed that for all temperatures, the $\qhat/T^3$ coefficient exhibits a logarithmic increase as the momentum $p$ of the jet rises. This momentum dependence is consistent with the asymptotic behavior of $\qhat$ predicted by perturbative Quantum Chromodynamics (pQCD). To explain the systematic differences between the models, we examine the pQCD limit for the DQPM, represented by the red dashed lines in Figure \ref{fig:qhat_p}. In this limit, the following considerations are made:\\
$\bullet$ all partons are considered on-shell, neglecting their widths.\\
$\bullet$ the exchange parton has a Debye mass in case of gluon $(\mu_D^g)^2= \dfrac{8\alpha_S}{\pi}(N_c+N_f)T^2$ or quark $(M_D^q)^2= \dfrac{2 \alpha_S}{\pi} C_F T^2$, while the scattered partons are assumed to be massless;\\
$\bullet$ the DQPM coupling is fixed to $\alpha_S = 0.3$, consistent with the LBT \cite{He:2015pra} and BAMPS models \cite{Senzel:2020tgf,Uphoff:2014cba}.\\
$\bullet$ classical (Maxwell-Boltzmann) statistics is employed.

In the high-energy limit $E/T \gg 1$ with $\alpha_s = 0.3$, the $\qhat/T^3$ shows a logarithmic scaling of $\qhat/T^3 = \text{const} \ \mathrm{ln} \left(\frac{q_{\text{max}}^2}{4\mu_D^2} \right) = \text{const} \ \mathrm{ln} \left(\frac{E}{T} \right)$ with $q_{\text{max}} = 2.6E_{\text{jet}}T$. A similar asymptotic behavior can be observed for the energy loss as well. Due to the different Debye masses for the LBT and BAMPS models, the results for $\qhat/T^3$ differ significantly at high $E/T$.

As observed in Fig. \ref{fig:qhat_p}, for a quark jet (left plot), the DQPM exhibits significantly larger values of $\qhat/T^3$ compared to the pQCD results from BAMPS and LBT models at low temperatures. This is attributed to the rise of $\alpha_S$ near $T_C$. However, when replacing $\alpha_S \to \alpha_{jet}=0.3$ at the jet parton vertices, the DQPM result decreases for low $T$ and becomes even smaller than the pQCD  models BAMPS and LBT. Moreover, the DQPM result in the pQCD limit (discussed above) is identical to the result of the BAMPS model (since the same Debye mass has been used in our calculations) for all $E/T$ models at low $T$, too. For the gluon jet (right plot) the DQPM shows again a reasonable agreement with pQCD models - BAMPS and LBT - for  $\alpha_{jet}=0.3$ and for the pQCD limit cases.


\section{Summary}

In this study, we have investigated the energy loss experienced by high-energy jet partons as they undergo elastic scattering with off-shell quarks and gluons while traversing the strongly interacting quark-gluon plasma (sQGP). The non-perturbative properties of the sQGP are described within the effective dynamical quasiparticle model (DQPM), which interprets the lQCD results on the QGP thermodynamics in terms of thermodynamics of off-shell quasiparticles with ($T,\mu_B$)-dependent masses and widths and broad spectral functions.

As we approached the critical temperature $T_C$, we observed a substantial rise in $\qhat/T^3$, which can be attributed to the increase in the strong coupling $\alpha_S(T,\mu_B)$ as $T$ approaches $T_C$. However, when we substituted $\alpha_S$ with a fixed value of $0.3$ at the jet parton vertex, the temperature dependence of $\qhat/T^3$ weakened, particularly in the pQCD limit. Furthermore, we noticed that $\qhat$ exhibited a significant increase with the momentum of the jets, indicating a strong momentum dependence. In comparison to pQCD models \cite{Senzel:2020tgf}, our results indicate that the energy loss experienced by a jet parton in the non-perturbative QCD medium (characterized by the DQPM) is stronger \cite{Grishmanovskii:2022tpb} when compared to scattering with massless pQCD partons. Additionally, for large temperatures, our $\qhat$ results demonstrate qualitative agreement with pQCD results, lattice results (both pure SU(3) and (2+1)-flavor), as well as with estimates from phenomenological studies by the JET and JETSCAPE collaborations and the Color String Percolation Model. However, it is important to note that for a quantitative comparison with phenomenologically extracted $\qhat$ values obtained from fitting jet observables measured in heavy-ion experiments, we must also consider the radiative energy loss. This aspect will be addressed in an upcoming study. We note that, the first calculations of the radiative processes in thermal sQGP within the DQPM has been done recently in Ref. \cite{Grishmanovskii:2023gog}.

Thus, our study of jet transport coefficients shows a large sensitivity of the jet energy loss to the properties of the QCD medium: weakly interacting pQCD versus the strongly interacting non-perturbative QGP.


\acknowledgments
We acknowledge support by the Deutsche Forschungsgemeinschaft (DFG, German Research Foundation) through the grant CRC-TR 211 "Strong-interaction matter under extreme conditions" - project number 315477589 - TRR 211. I.G. also acknowledges support from the "Helmholtz Graduate School for Heavy Ion research". 


\bibliographystyle{JHEP}
\bibliography{refs}

\end{document}